\documentclass[aps,prd,twocolumn,showpacs,groupedaddress]{revtex4}
\usepackage{graphicx}
\usepackage{latexsym}
\usepackage{subfigure}

\begin{document}
\title{Determination of the Neutrino Mass Hierarchy at an Intermediate Baseline}

\author{Liang Zhan, Yifang Wang, Jun Cao, Liangjian Wen}
%\email{zhanl@ihep.ac.cn}

\affiliation{Institute of High Energy Physics, Beijing, 100049}

\begin{abstract}
It is generally believed that neutrino mass hierarchy can be
determined at a long baseline experiment, often using accelerator
neutrino beams. Reactor neutrino experiments at an intermediate
baseline have the capability to distinguish normal or inverted
hierarchy. Recently it has been demonstrated that the mass hierarchy
could possibly be identified using Fourier transform to the L/E
spectrum if the mixing angle $\sin^2(2\theta_{13})>0.02$. In this
study a more sensitive Fourier analysis is introduced. We found that
an ideal detector at an intermediate baseline ($\sim 60$~km) could
identify the mass hierarchy for a mixing angle $\sin^2(2\theta_{13})
> 0.005$, without requirements on accurate information of
reactor neutrino spectra and the value of $\Delta m^2_{32}$.
\end{abstract}

\maketitle

Recent results  from solar, atmospheric, reactor and accelerator
neutrino experiments all show that neutrinos are massive and they
can oscillate from one type to another. Among all the six mixing
parameters, three of them are known, two unknowns, and one of them,
the mass-squared difference $\Delta m^2_{32}$, defined as $m^2_3 -
m^2_2$, is only known to be $ |\Delta m^2_{32}|= (2.43\pm 0.13)
\times 10^{-3} {\rm eV}^2$ ($68\%$ C.L.) from accelerator neutrino
experiments \cite{Adamson:2008zt}.  The question, if the mass
hierarchy is normal ($\Delta m^2_{32}>0$) or inverted ($\Delta
m^2_{32}<0$), is not known now but is fundamental to particle
physics.

For normal hierarchy (NH) or inverted hierarchy (IH),  the
 neutrino mass-squared difference has the following relations:
\begin{eqnarray}
\label{eq:masssqare}
 \Delta{m}^2_{31} &=& \Delta{m}^2_{32}+\Delta{m}^2_{21} \nonumber \\
 {\rm NH:~} |\Delta{m}^2_{31}| &=&
 |\Delta{m}^2_{32}|+|\Delta{m}^2_{21}| \nonumber \\
  {\rm IH:~} |\Delta{m}^2_{31}| &=&
 |\Delta{m}^2_{32}|-|\Delta{m}^2_{21}|
\end{eqnarray}
In principle, the mass hierarchy can be determined by precision
measurements of $|\Delta{m}^2_{31}|$ and $|\Delta{m}^2_{32}|$. In
fact  it is extremely difficult since $\Delta m^2_{21}$ is only
$\sim3\%$ of $|\Delta{m}^2_{32}|$, hence $|\Delta{m}^2_{32}|$ and
$|\Delta{m}^2_{31}|$ have to be measured with a precision much
better than 3~\%.

Effects of mass hierarchy can be amplified by matter effects if the
baseline is large enough, say several hundreds to thousands of
kilometers. Such experiments often need accelerator-based neutrino
beams and huge detectors. Proposals such as T2K \cite{Itow:2001ee,
Mena:2006uw}, Nova \cite{Mena:2005ri, Ayres:2004js, Mena:2006uw} and
T2KK \cite{Hagiwara:2006vn} have mass hierarchy sensitivity in the
$\nu_{\mu}\rightarrow \nu_e$ channel if $\theta_{13}$ is large
enough ($i.e.\sin^2(2\theta_{13})\geq 0.03$). In addition, they are
affected by the ($\delta_{CP}$, $sign(\Delta m^2_{32})$) degeneracy
\cite{Minakata:2001qm, Barger:2001yr}. At a magic baseline
\cite{Huber:2003ak, Smirnov:2006sm}, $L\sim 7000~km$, the degeneracy
can be canceled but it requires a very intensive source such as a
neutrino factory or a beta-beam which will not be available in the
near future. A method using atmospheric neutrinos
\cite{Gandhi:2008fu, Samanta:2006sj} with a baseline of
$L\sim10^4~km$ and the neutrino energy of $E\sim 1~GeV$ is sensitive
to mass hierarchy for very small or even null value of
$\theta_{13}$, if the measurement precision of $|\Delta{m}^2_{32}|$
is better than 2\%.

Method using reactor neutrino based intermediate baseline
($40-65~km$) experiments has been explored based on precision
measurement of distortions of the energy spectrum due to non-zero
$\theta_{13}$ \cite{Petcov:2001sy, Choubey:2003qx}. Recently, a
study \cite{Learned:2006wy} shows a new method to distinguish normal
or inverted hierarchy after a Fourier transform of the L/E spectrum
of reactor neutrinos. It is observed that the Fourier power spectrum
has a small shoulder next to the main peak, and their relative
position can be used to determine the mass hierarchy.  A¡°filter
method¡± is used to improve the sensitivity to the mass hierarchy up
to $\sin^2(2\theta_{13})>0.02$, if $\Delta m^2_{32}$ is known $a~
priori$. Comparing to a normal L/E analysis, the Fourier analysis
naturally separates the mass hierarchy information from
uncertainties of the reactor neutrino spectra and other mixing
parameters, which is critical for very small $\sin^2(2\theta_{13})$
oscillations.

In this paper, we report that if a proper Fourier transform is
applied and if all information is fully utilized, the capability of
an intermediate baseline reactor experiment to determine the
neutrino mass hierarchy can be improved for a smaller mixing angle
$\theta_{13}$ without knowing $\Delta m^2_{32}$ $a ~priori$. In the
following, we will use a reactor neutrino spectrum to illustrate the
method, but such a method can be generalized to other experiments.

For a reactor neutrino experiment, the observed neutrino spectrum at a baseline L,
$F(L/E)$, can be written as
$$ F(L/E) = \phi(E)\sigma(E) P_{ee}(L/E) $$
where $E$ is the electron antineutrino ($\overline{\nu}_e$) energy,
$\phi(E)$ is the flux of $\overline{\nu}_e$ from the reactor,
$\sigma(E)$ is the interaction cross section of $\overline{\nu}_e$
with matter, and $P_{ee}(L/E)$ is the $\overline{\nu}_e$ survival
probability.

The $\overline{\nu}_e$ flux  $\phi(E)$ from the reactor can be
parameterized as \cite{Vogel:1989iv},
\begin{eqnarray}
\label{eq:flux}
 \phi(E) &=&0.58{\rm Exp}(0.870 - 0.160E - 0.091E^{2})\nonumber\\
 &+&0.30{\rm Exp}(0.896 - 0.239E - 0.0981E^{2})\nonumber\\
 &+&0.07{\rm Exp}(0.976 - 0.162E - 0.0790E^{2})\nonumber\\
 &+&0.05{\rm Exp}(0.793 - 0.080E - 0.1085E^{2}),
\end{eqnarray}
where four exponential terms are contributions from isotopes
$^{235}$U, $^{239}$Pu, $^{238}$U and $^{241}$Pu in the reactor fuel,
respectively.

The leading-order expression for the cross section
\cite{Vogel:1999zy} of inverse-$\beta$ decay ( $\overline{\nu}_e +
p\rightarrow e^+ + n$ ) is
\begin{eqnarray}
\label{eq:crosssection}
 \sigma^{(0)} = 0.0952 \times 10^{-42} {\rm cm}^2
(E^{(0)}_e p^{(0)}_e/1 {\rm MeV^2})
\end{eqnarray}
where $E^{(0)}_e = E_\nu - (M_n - M_p)$ is the positron energy when
neutron recoil energy is neglected, and $p^{(0)}_e$ is the positron
momentum. The survival probability of $\overline{\nu}_e$ can be
expressed as \cite{Bilenky:2001jq}
\begin{eqnarray}
\label{eq:Pee}
 P_{ee}(L/E) &=& 1 - P_{21} - P_{31} - P_{32} \nonumber \\
 P_{21} &=& \cos^4(\theta_{13})\sin^2(2\theta_{12})\sin^2(\Delta_{21})\nonumber\\
 P_{31} &=& \cos^2(\theta_{12})\sin^2(2\theta_{13})\sin^2(\Delta_{31})\nonumber\\
 P_{32} &=& \sin^2(\theta_{12})\sin^2(2\theta_{13})\sin^2(\Delta_{32})
\end{eqnarray}
where $\Delta_{ij} = 1.27 \Delta m^{2}_{ij}L/E$, $\Delta m^{2}_{ij}$
is the neutrino mass-squared difference ($m^2_i - m^2_j$) in eV$^2$,
$\theta_{ij}$ is the neutrino mixing angle,  $L$ is the baseline
from reactor to $\overline{\nu}_e$ detector in meters, and $E$ is
the $\overline{\nu}_e$ energy in MeV.

$P_{ee}(L/E)$ has three oscillation components, $P_{21}$, $P_{31}$
and $P_{32}$, corresponding to three oscillation frequencies in
$L/E$ space, which are proportional to $|\Delta m^2_{ij}|$,
respectively. Their relative amplitude(oscillation intensity), is
about $40:2:1$ from a global fit \cite{Maltoni:2004ei} of mixing
parameters as listed in Table {\ref{tab:globalfit}}. The oscillation
component $1 - P_{21}$ dominates the $P_{ee}$ oscillation, while
$P_{31}$ and $P_{32}$, which are sensitive to the neutrino mass
hierarchy, are suppressed by the small value of
$\sin^2(2\theta_{13})$.
\begin{table}
\begin{tabular}{|c|c|c|c|}
  \hline
  parameter & ~best fit~ & 2$\sigma$ & 3$\sigma$  \\
  \hline
  ~$\Delta m^2_{21} [10^{-5}{\rm eV}^2]$~ & 7.6 & 7.3-8.1 & 7.1-8.3  \\
  ~$|\Delta m^2_{32}| [10^{-3}{\rm eV}^2]$~ & 2.4 & 2.1-2.7 & 2.0-2.8  \\
  $\sin^2\theta_{12}$ & 0.32 & ~0.28-0.37~ & ~0.26-0.40~ \\
  $\sin^2\theta_{23}$ & 0.50 & 0.38-0.63 & 0.34-0.67 \\
  $\sin^2\theta_{13}$ & 0.007 & $\leq 0.033$ & $\leq 0.050$ \\
  \hline
\end{tabular}
\caption{Neutrino mixing parameters from a global fit, updated in
2007, as the inputs to this study.} \label{tab:globalfit}
\end{table}

\begin{figure}[htbp]
\begin{center}
\includegraphics[width=0.45\textwidth]{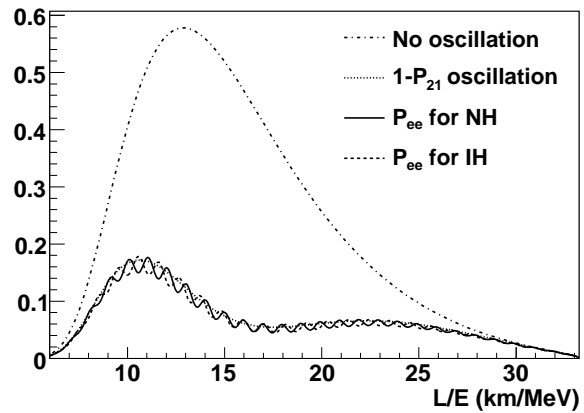}
\caption{ Reactor neutrino spectra at a baseline of 60 km in L/E
space for no oscillation (dashed dotted line), $1-P_{21}$
oscillation (dotted line) and $P_{ee}$ oscillation in the cases of
NH and IH, assuming $\sin^2(2\theta_{13})=0.1$.} \label{fig:nuEspec}
\end{center}
\end{figure}

The observed neutrino spectrum in L/E space, taking the baseline L
to be 60~km and all the other parameters from Table
{\ref{tab:globalfit}} except $\sin^2(2\theta_{13})$, is shown in
Fig.{\ref{fig:nuEspec}}, together with that of no oscillation. For
comparison, the oscillation spectrum without $P_{31}$ and $P_{32}$
are also shown. For a very small $\sin^2(2\theta_{13})$, a normal
$\chi^2$ analysis on the L/E spectrum with binned data, which
requires accurate knowledge on the neutrino energy spectra and much
smaller binning than the energy resolution, is difficult for the
mass hierarchy study.

Since neutrino masses all appear in the frequency domain as shown in
Eq.~\ref{eq:Pee}, a Fourier transform of $F(L/E)$ shall enhance the
sensitivity to the mass hierarchy. The frequency spectrum can be
obtained by the following Fourier sine transform(FST) and Fourier
cosine transform(FCT):
\begin{eqnarray}
\label{eq:FT}
FST(\omega) = \int^{t_{max}}_{t_{min}}F(t) \sin (\omega t)\mathrm{d}t \nonumber \\
FCT(\omega) = \int^{t_{max}}_{t_{min}}F(t) \cos(\omega
t)\mathrm{d}t
\end{eqnarray}
where $\omega$ is the frequency, $\omega=2.54 \Delta m^{2}_{ij}$;
$t=\frac{L}{E}$ is the variable in L/E space, varying from $t_{min}
= \frac{L}{E_{max}}$ to $t_{max} = \frac{L}{E_{min}}$.

Since $P_{ee}$ is a linear combination of $1-P_{21}$, $P_{31}$ and
$P_{32}$, FST and FCT spectra can be divided into three components
corresponding to $1-P_{21}$, $P_{31}$ and $P_{32}$ respectively.
Fig.{\ref{fig:ftSpec}} shows the three components of the FST and FCT
spectra together with full $P_{ee}$ oscillation for both NH and IH
cases. The oscillation frequency is proportional to $\Delta
m^2_{ij}$, so we can scale the frequency to be $\delta m^2$ and plot
the spectra in axis of $\delta m^2$ in the interested frequency
range of $1.8\times 10^{-3} {\rm eV^2}<\delta m^2<3.0\times 10^{-3}
{\rm eV^2}$. From Fig.{\ref{fig:ftSpec}}, we know that:

\begin{figure}[!htb]
\centering
\includegraphics[width=0.48\textwidth]{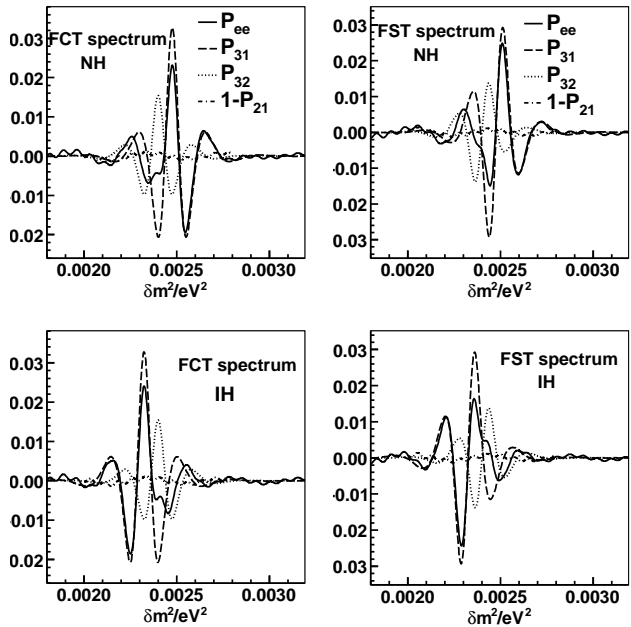}
\caption{Fourier sine (FST) and cosine (FCT) transform spectra for
$1-P_{21}$ component (dotted line), $P_{32}$ component (dashed
line), $P_{31}$ component (dot-dashed line) and all the components
of $P_{ee}$ (solid line) in the cases of NH and IH. }
\label{fig:ftSpec}
\end{figure}

\begin{enumerate}
    \item  $P_{31}$ and $P_{32}$ components dominate the FCT and FST
    spectra in the interested frequency range of $1.8\times 10^{-3} {\rm
    eV^2}<\delta m^2<3.0\times 10^{-3} {\rm eV^2}$ since
     $|\Delta m^2_{31}|$ and $|\Delta m^2_{32}|$
     are in this range, while $1-P_{21}$ is very weak since its
    oscillation frequency is in a much lower
    range. The FST and FCT spectra of $P_{ee}$
    are approximately the sum of $P_{31}$ and $P_{32}$ components which
     are sensitive to mass hierarchy.

    \item For NH, the $P_{32}$ FCT and FST spectra are left-shifted
    with respect to the $P_{31}$ spectra because $|\Delta{m}^2_{32}| <
 |\Delta{m}^2_{31}|$; while for IN, the $P_{32}$ spectra are
 right-shifted because $|\Delta{m}^2_{32}| >
 |\Delta{m}^2_{31}|$.

    \item The peak of FCT spectrum corresponds to
    the zero point of FST spectrum. This feature is helpful to
    identify the position of $|\Delta m^2_{32}|$ and $|\Delta
    m^2_{31}|$, without knowing their accurate values $a~priori$.

    \item For FCT spectrum, $P_{32}$ and
    $P_{31}$ components have similar shapes with the peak around
    $|\Delta m^2_{32}|$ and $|\Delta m^2_{31}|$, respectively, and two valleys on
    each side of the peak. The amplitude of $P_{32}$ to that of
    $P_{31}$ has a ratio of about 1:2 determined by
    tan$^2(\theta_{12})$. The shapes of $P_{32}$ and $P_{31}$
    are left-right symmetric with respect to their peaks (mirror symmetric).
     This symmetry is broken for $P_{ee}$ as an
     approximate sum of $P_{32}$ and $P_{31}$ in different ways for
     NH and IH. For NH, the peak of $P_{32}$ is
    at the left of the valley of $P_{31}$, while for
    IH, the peak of $P_{32}$ is at the right of the valley of $P_{31}$. This
    feature can be used to distinguish NH and IH.

    \item For FST spectrum, the shapes of $P_{32}$ and $P_{31}$
     are positive-negative symmetric with respect to zero (rotation symmetric)
     around $|\Delta m^2_{32}|$ and $|\Delta m^2_{31}|$, respectively.
     This symmetry is broken for $P_{ee}$ in different ways for NH and IH.
     For NH, the peak of $P_{32}$ is at the valley position of $P_{31}$,
     while for IH, the valley of $P_{32}$ is at the peak position of $P_{31}$.
     This feature can be also used to distinguish NH and IH.

\end{enumerate}

As discussed above and shown in Fig.~\ref{fig:ftSpec}, the normal or
inverted mass hierarchy can be distinguished by the symmetry
breaking features of the FCT and FST spectra. To quantify these
features, two parameters, RL and PV, are introduced as the
following:
\begin{eqnarray}
\label{eq:defRLPV}
 RL = \frac{RV-LV}{RV+LV}\,,~ PV = \frac{P-V}{P+V}\,,
\end{eqnarray}
where RV is the amplitude of the right valley and LV is the
amplitude of the left valley in the FCT spectrum. P is the amplitude
of the peak and V is the amplitude of the valley in the FST
spectrum. From the above discussion, we know
\begin{eqnarray}
\label{eq:RLPVcondition}
 RL > 0 ~&{\rm and}& PV > 0 ~\Rightarrow ~{\rm NH} \nonumber \\
 RL < 0 ~&{\rm and}& PV < 0 ~\Rightarrow ~{\rm IH}
\end{eqnarray}

The values of $RL$ and $PV$ as well as the shapes of FCT and FST
spectra depend on the baseline and neutrino mixing parameters.
Parameters such as $\sin^2\theta_{12}$, $\Delta m^2_{21}$, and
$\Delta m^2_{32}$ are relatively well known, hence only small
uncertainties are introduced.  The baseline and
$\sin^2(2\theta_{13})$ are more important and are discussed below.
\begin{enumerate}
    \item Baseline determines the oscillation cycles. To maximize the
    symmetry breaking of FCT and FST spectra, we scan the baseline length and
    find that the peak (valley)of $P_{32}$ spectrum lays on the valley (peak)
    of $P_{31}$ spectrum around 60~km.
    The widthes of peaks and valleys of the Fourier spectra,
    which are proportional to 1/L, are also determined by baseline. In an extreme case, the peak and valley of
    $P_{31}$ and $P_{32}$ spectra all become $\delta$-functions at
    infinite baseline, hence are well separated from each other.
    In fact, this is already the case at 200~km and
    the mass hierarchy can be determined by looking at the position of the smaller
    peak ($P_{32}$ component). If it is on the left side of the main peak ($P_{31}$ component), it is NH.
    Otherwise it is IH. However, since the neutrino flux from reactors is proportional to $1/L^2$,
    shorter baseline, say at 60 km, is the best from an experimental point of view. The actual optimum
   baseline can be determined by taking into account both statistical and systematical errors.

    \item $\sin^2(2\theta_{13})$ determines the amplitude of the Fourier spectra of $P_{31}$ and
    $P_{32}$. At $\sin^2(2\theta_{13}) = 0$, $P_{31}$ and $P_{32}$ components will
    vanish and no features can be used to discriminate the mass hierarchy. A minimum value of
   $\sin^2(2\theta_{13})$ to distinguish NH and IH experimentally will be
   analyzed by taking into account possible experimental errors ~\cite{exp}.
\end{enumerate}

\begin{figure}[!htb]
\centering
\includegraphics[width=0.40\textwidth]{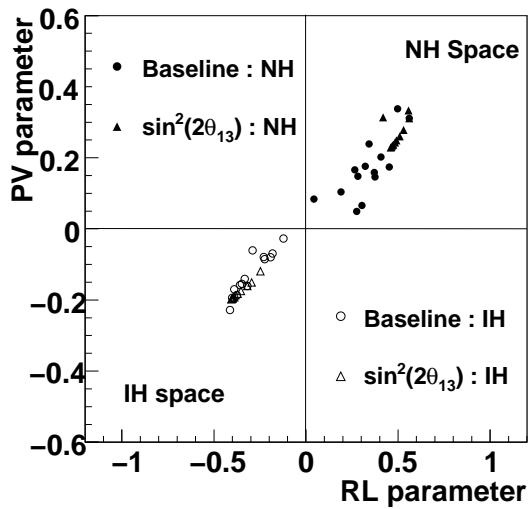}
\caption{Distribution of RL and PV values for different parameters
of baseline and $\sin^2(2\theta_{13})$. For each parameter to be
scanned, the default baseline is 60 km and all the other parameters
are the values as in Table~{\ref{tab:globalfit}}. Two clusters of RL
and PV values are clearly seen for NH and IH cases.}
\label{fig:parScan}
\end{figure}

In order to understand the robustness of the discrimination method
using FCT and FST spectra,  values of baseline are scanned from 46
to 72~km; $\sin^2(2\theta_{13})$ from 0.005 to 0.05. The resultant
RL and PV values are well separated into two clusters, corresponding
to the case of NH and IH respectively, as shown in
Fig.{\ref{fig:parScan}}.

\begin{figure}[!htb]
\centering
\includegraphics[width=0.40\textwidth]{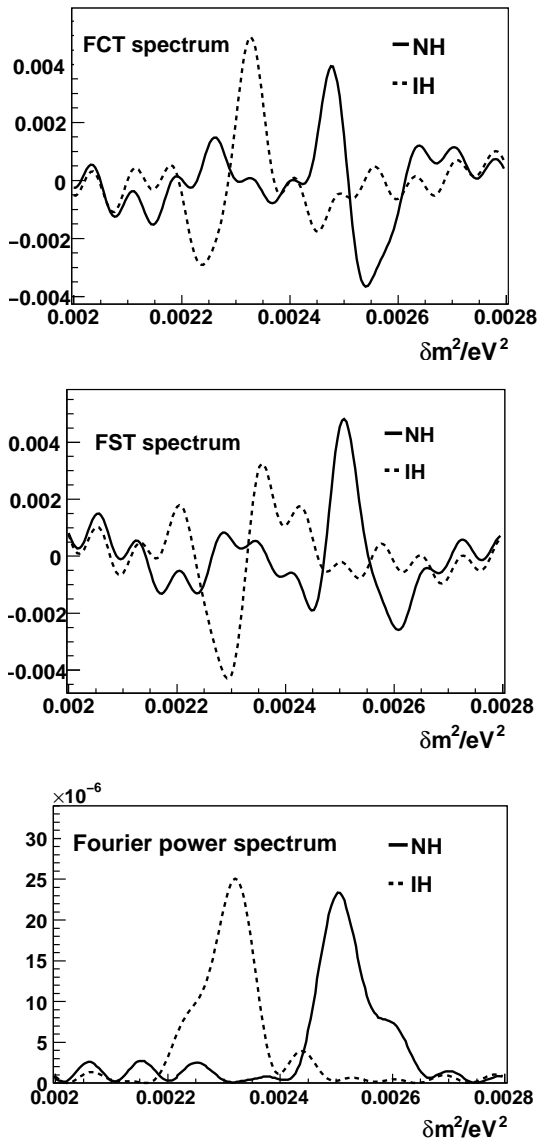}
\caption{The FCT and FST spectra and Fourier power spectrum for
 $\sin^2(2\theta_{13}) = 0.005$. The solid line is for NH and the
 dashed line is for IH. The FCT and FST spectra have distinctive features
 to identify the mass hierarchy, which looks more sensitive than the Fourier power spectrum method.}
\label{fig:FT} % caption for the whole figure
\end{figure}

The FCT and FST spectra for $\sin^2(2\theta_{13}) = 0.005$ are shown
in Fig.{\ref{fig:FT}}. Although a detailed experimental analysis of
error contour is to be completed~\cite{exp}, the features of NH and
IH are still very distinctive. On the FCT spectrum, a valley appears
at the left of the prominent peak for IH, and a peak appears at the
left of the valley for NH. On the FST spectrum, there is a clear
valley for IH, while for NH it is a peak. In comparison, the Fourier
power spectrum used in Ref.~\cite{Learned:2006wy} is also shown in
Fig.{\ref{fig:FT}}. The FCT and FST method is more sensitive than
the Fourier power spectrum method for a very small
$\sin^2(2\theta_{13})$.

For even smaller $\sin^2(2\theta_{13})$, the main peak becomes less
significant. For example, if the main peak is required to be twice
higher than that of noise, $\sin^2(2\theta_{13})$ must be greater
than 0.005 in order to clearly identify the main peak, for a variety
of neutrino energy spectra
 in a reasonable range.

For a realistic experiment in the near future, the energy resolution
and statistics are of the most concern. At 60~km, $\theta_{12}$ has
the least impact to the mass hierarchy determination. The energy
resolution must be good enough not to smear the difference between
$P_{31}$ and $P_{32}$, which requires the energy resolution be
better than $3\%/\sqrt{E}$. A detector with a mass at 10 kton level
may be necessary, depending on the size of $\theta_{13}$. If
shortening the baseline, the noise in the Fourier spectra from
$\theta_{12}$ oscillation increases, thus degrade the sensitivity.
In the mean time requirements to the energy resolution and the
detector size  are relaxed. The optimization of the baseline as well
as the energy resolution and detector size for different
$\theta_{13}$ assumptions are undergoing.

In summary, the method to discriminate the mass hierarchy has been
studied by using a Fourier sine(FST) and cosine(FCT) transform to
the observed reactor neutrino L/E spectra. The FCT and FST spectra
can separate $P_{31}$ and $P_{32}$ oscillation components from the
large $1-P_{21}$ component in a specific $\delta m^2$ range.
Features of mass hierarchy are enhanced in this representation and
more sensitive than that of the Fourier power spectrum at very small
$\sin^2(2\theta_{13})$. We found that an ideal detector at an
intermediate baseline ($\sim 60$~km) could identify the mass
hierarchy for a mixing angle $\sin^2(2\theta_{13}) > 0.005$, without
requirements on accurate information of reactor neutrino spectra and
the value of $\Delta m^2_{32}$. A paper of a detailed analysis of
experimental errors will be released soon~\cite{exp}. Similar
methods can be applied to other experiments using different neutrino
sources, such as accelerator-based neutrino beams or atmospheric
neutrinos.

\end{document}